\documentclass[12pt]{article}

\usepackage{amsmath,amsfonts,amssymb,latexsym,graphicx}

\numberwithin{equation}{section}
\def\be{\begin{equation}}
\def\ee{\end{equation}}
\def\bq{\begin{eqnarray}}
\def\eq{\end{eqnarray}}
\def\beq{\begin{eqnarray}}
\def\eeq{\end{eqnarray}}

\def\a{\alpha}
\def\b{\beta}

\def\m{\mu}

\begin{document}
\title{\textsc{The conformal cosmological potential}}
\author{Spiros Cotsakis$^{1,2}$\thanks{\texttt{skot@aegean.gr}},
Ifigeneia Klaoudatou$^2$\thanks{\texttt{iklaoud@aegean.gr}},
Georgios Kolionis$^2$\thanks{\texttt{gkolionis@aegean.gr}},\\
John Miritzis$^3$\thanks{\texttt{imyr@aegean.gr}},
Dimitrios Trachilis$^4$\thanks{\texttt{Dimitrios.Trachilis@aum.edu.kw}}
\\[10pt]
$^{1}$Institute of Gravitation and Cosmology, RUDN University\\
ul. Miklukho-Maklaya 6, Moscow 117198, Russia
\\ $^{2}$Research Laboratory of Geometry,  Dynamical Systems\\  and Cosmology,
University of the Aegean,\\ Karlovassi 83200,  Samos, Greece
\\ $^3$Department of Marine Science,
University of the Aegean,\\
Mytilene, Greece
\\ $^4$College of Engineering and Technology,\\
American University of the Middle East, Kuwait}
\date{March 2022}
\maketitle
\newpage
\begin{abstract}
\noindent
We discuss qualitative features of the conformal relation between certain classes of gravity theories and general relativity, common to different themes such as $f(R)$, Brans-Dicke-type, and string theories. We focus primarily on the frame relations of the fields involved, slice energy, traceless and Palatini extensions, and selected cosmological applications.
\end{abstract}
\newpage
\tableofcontents
\newpage
\section{Introduction}
Relativistic causality, that is deciding which events can influence or be influenced by other events, is a fundamental attribute  of the geometry and dynamics of spacetime. In general relativity (GR), causality appears together with the imposition of the Einstein equations on spacetime, and their interplay leads to many distinctive results, such as the singularity theorems of Hawking and Penrose and the black hole area theorem~\cite{witten0}.

A wider appreciation of the restrictions on spacetime placed by the imposition of relativistic causality  can be developed by studying geometric extensions of Einstein's theory,  where, while the causality principle is respected, the solution space of the field equations is distinctively different than in GR. 
The only transformations that respect the null structure (and therefore give identical local causal geometry in a neighbourhood of any event) are of course the conformal maps from a given metric $g_{ab}$ to a conformally related one $\tilde{g}_{ab}$,  obtained by a multiplication of the original metric by a non-zero function, the conformal factor $\Omega$ defined on spacetime,
\begin{equation} \label{conf}
    \tilde{g}_{ab}=\Omega^{2}g_{ab}.
\end{equation}

The Einstein equations are of course not conformally invariant and one is therefore led to the consideration of a broader program, not restricted to  `just' the Einstein equations for a conformal metric (a problem already deep enough and which we do not discuss here), dealing with the development and  demarcation of the possible consequences  of studying equations  roughly `within the conformal envelope of GR', that is, theories which in the broadest sense include gravity and at the same time  are conformally \emph{related} to GR, and thus share the same causal (null) structure as the latter. Some basic aspects of this program are discussed below.

This program is a very ambitious one and admits to its study any set of field equations conformally related to the Einstein field equations for the metric $g_{ab}$ and possible matter fields $\psi$. Since a conformal map is  transitive with respect to composition, the richness of the class of conformally related but otherwise distinct theories of gravity is truly bewildering. Presently, however, we shall mostly be interested in the following three classes of such theories:
\begin{enumerate}
\item $f(R)$ theory
\item Brans-Dicke-type theories
\item Effective theories of a quantum/string origin.
\end{enumerate}

Although all three  types are related to each other (most notably by duality transformations), and can also be further restricted or generalized by excluding or including further possibly self-interacting fields, they share the common property of having a \emph{direct} conformal relation to Einstein's theory (generally valid in higher than four dimensions, and sometimes after compactification).

In this paper, we review some results in this program proven useful mainly for cosmological considerations, by focusing mostly on the first variant. Generally speaking, the  results reviewed here on $f(R)$ theory are also broadly indicative of the behaviour of the other two types of theory (in the case of string actions this is true for only  the gravi-dilaton part of the full theory-see below). The conformal relation of the Brans-Dicke theory to GR was discovered in 1962 by Robert Dicke~\cite{d2}, while that of the various string actions is carefully  developed in~\cite{gasp}, chaps. 2, 3.

While the standard Einstein-Hilbert lagrangian density is proportional to  $R$, $f(R)$-theory is  based on a $D$-dimensional action proportional to the lagrangian density,
\begin{equation} \label{fRLagran}
   L_{f(R)}=f(R),
\end{equation}
where $f(R)$ is an analytic function of the scalar curvature $R$ of the metric $g_{ab}$. All five superstring lagrangian densities in $D$ spacetime dimensions contain the same basic `gravi-dilaton' part described by the lagrangian density,
\be \label{string}
L_{\textrm{string}}=e^{-\phi}(R+(\nabla\phi)^2),
\ee
which by the scalar field redefinition $e^{-\phi}=\Phi$ is identical to the $\omega=-1$ Brans-Dicke lagrangian density,
\be \label{BD}
L_{\textrm{BD}}=\Phi R -\omega\Phi^{-1} (\nabla\Phi)^2,
\ee
with $\omega$ being the usual Brans-Dicke parameter. However, all five different string actions (i.e., those of type I, IIA, IIB, heterotic type-I, and type-II) containing the part~(\ref{string}), also contain extra fields with different couplings to the dilaton (and are further generalized by including an arbitrary dilaton potential). The sixth action of interest, the so-called 11-dimensional supergravity theory, is a different one, but becomes a type IIA theory of the form~(\ref{string}) (with extra fields) after compactification on a circle, with the scalar field $\phi$ arising in this case as the dynamical scale factor of the 11th dimension.

All three classes of theory~(\ref{fRLagran})--(\ref{BD}) are in general augmented by the addition of the Gibbons-Hawking boundary term as well as a  matter Lagrangian corresponding to all other fields. The use of effective action techniques to describe early universe phenomena is critically discussed in~\cite{bra1}.

The vacuum field equations  that arise by varying the associated action with respect to the metric $g_{ab}$ are
\begin{equation}\label{fRVeqns}
    f'R_{ab}-\frac{1}{2}fg_{ab}-\nabla_{a}\nabla_{b}f'+g_{ab}\Box f'=0,
\end{equation}
with $f'$ denoting the derivative of the function $f(R)$ with respect to the scalar curvature $R$, and $\Box\equiv g^{ab}\nabla_{a}\nabla_{b}$. It is a well known result that the $f(R)$ Equation~(\ref{fRVeqns}) are conformally related to the Einstein equations with a self-interacting scalar field which has a particular potential.
The conformal equivalence theorem then states that under the conformal transformation~(\ref{conf}) and upon introducing,
\begin{equation}\label{phidefn}
    \Omega^2=e^\phi=f'(R),
\end{equation}
the field equations~(\ref{fRVeqns}) become those  of general relativity with a scalar field matter source,
\begin{equation}\label{einfr}
    \tilde{G}_{ab}=\tilde{T}_{ab}(\phi),
\end{equation}
where $\tilde{G}_{ab}$ is the Einstein tensor, $\tilde{T}_{ab}(\phi)$ is the stress-energy tensor of the scalar field $\phi$, and  the scalar field has an effective potential of  the form~\cite{B-C88},
\begin{equation}\label{pot}
    V(\phi)=\frac{1}{2}(f')^{D/(2-D)}(Rf'-f).
\end{equation}

In this work we focus  on certain aspects exclusively connected with this conformal relation, but do not cover much of the applications that, perhaps surprisingly, stem from this relation: the possibility of inflation in both the original and the conformal frames  (which is developed already in numerous early papers on this subject, e.g.,~\cite{star79,ma88,Bar88,B-B88,Hall87,C-S93,A-W84,L-M85}), the existence and nature  of spacetime singularities, the asymptotic behaviour (both to the past and the future) of various types of cosmological models, and many others.

This  paper splits into two parts, with the first part comprising Sections \ref{confthm}--\ref{sec5} dealing with foundational matters, while the last three Sections form the second part and deal with certain further issues important for cosmological applications. There are many more aspects of these theories than we  cover here, and the interested reader is referred to the reviews~\cite{copeland,clifton,capo1,od1,od2,eff}, while below we give a few extra references to some more specialized topics.

\section{Conformal Equivalence}\label{confthm}
The conformal equivalence theorem dictates that the higher-order $f(R)$ gravitational equations (or any of the other two classes of theory mentioned in the Introduction) for the Jordan frame metric $g_{ab}$ in vacuum become conformally the Einstein equations for the comformal metric $\tilde{g}_{ab}$ with a nontrivial  scalar field $\phi$ as source in the Einstein frame. We shall use the phrases \emph{Jordan} (or `original') \emph{frame}, and \emph{Einstein} (or `conformal') \emph{frame} to imply the field equations and conditions written in terms of the original metric $g_{ab}$ and correspondingly in terms of the conformal metric $\tilde{g}_{ab}$. If matter is added in either the original  or the conformal frame, extra terms containing a $\phi$-matter interaction generally appear.

A few pertinent comments about the conformal equivalence theorem are now given. Firstly, there are two aspects of the meaning of the word `equivalence' in the theorem, the mathematical and the physical. Mathematically,
the equivalence between {\hbox {Equations~(\ref{fRVeqns}) and~(\ref{einfr})}} is effective at all points on the spacetime $M$ such that $\Omega\neq 0$, or $\infty$.  In other words,  when the conformal map~(\ref{conf}) is non-singular, the conformal relation between the two sets of field equations appears, and we can carry freely between the two conformally related spacetimes $(M,g), (M,\tilde{g})$ all causal structure results, for instance, causality conditions, achronality, Cauchy developments, etc., but not those that use or are implied by the field equations, like, for instance,  the energy conditions.

Secondly, the scalar function $\phi$ introduced above as equal to $\ln f'$ describes possible changes in the gravitational Lagrangian $f(R)$ when the scalar curvature $R$ changes from point to point in spacetime. The field $\phi$ has nothing to do, and cannot be identified  with, other scalar fields usually considered in the literature, for example,  a scalar field coming from electroweak theory such as the Higgs field. This `$f(R)$ scalar field'  is by its definition very slowly changing with $f'$ (defined as $\propto\ln f'$) and therefore of a different nature than other particle physics-based fields. The scalar field  $\phi$,  is perhaps closer in spirit to the Brans-Dicke theory~\cite{d1,d2}, or to the dilaton field present in the string cosmology equations. By its definition, $\phi$ is interpreted as part of the gravitational field in the Jordan frame, and after the conformal transformation it is considered as a `matter field', with the total    `matter' lagrangian density being $L_{\textrm{mat}}+L_{\phi}$.

Thirdly, this mathematical correspondence has, however,  important interpretational implications for the physical importance of the conformal transformation. In general, physical processes distinguish the Jordan frame $f(R)$ (or scalar-tensor, like the Brans-Dicke) theories from other particle physics models taking place in the Jordan frame, as well as from those in the Einstein frame. This is generally  true in the context of $f(R)$ and Brans-Dicke theories with minimal coupled matter fields as well as for the bosonic forms in the RR sector of type IIA and IIB superstring theories. String theory phenomenology predicts a non-minimal and non-universal coupling of the matter fields to the `dilaton' $\phi$, cf.~\cite{gasp}. As another kind of  example, since inflation is admitted in such theories~\cite{B-C88}, it was suspected very early on that the usual process of thermalization considered  in  inflationary theory driven by a phase transition in the scalar field dynamics is not applicable here, and  instead, it would be possible that reheating is successful without a phase transition using $f(R)$ theory. This has been demonstrated very early on for special cases (see for instance the case of a quadratic lagrangian with a flat FRW metric~\cite{mms}).

Fourth, the cosmic field  $\phi$ has a potential energy given by Equation~(\ref{pot}) in the vacuum $f(R)$ theory. This potential energy  measures the failure of the gravitational field to be described by the vacuum Einstein equations of GR, since it equals zero when the gravitational lagrangian assumes its Einstein-Hilbert form without matter. That is, the scalar field potential energy is exactly zero for GR. Therefore, a future observational measurement of a non-zero $V$  given by Equation~(\ref{pot}) would point to the existence of $f(R)$  higher-order theories of gravity, or of string cosmological effects, cf.~\cite{gasp}, chap. 9.

We now turn our attention to theories of gravity with many  scalar fields that couple arbitrarily to themselves and to the curvature given by the action,
\be\label{wm1}
S=\int_{\mathcal{M}}\sqrt{-g}d^{m} x\left(  A(\phi)R-B(\phi)g^{\mu\nu}
h_{ab}\partial_{\mu}\phi^{a}\partial_{\nu}\phi^{b}\right)  ,
\ee

We call such an action a \emph{wavemap-tensor theory}.
A wavemap is a map $\phi: \mathcal{M}\longrightarrow \mathcal{N}$ from a spacetime manifold $(\mathcal{M}^{m},g_{\mu\nu})$-the source manifold, to any (semi-)Riemannian manifold $(\mathcal{N}^{n},h_{ab})$, the target manifold, described in the target manifold by its coordinates $(\phi^a)$, $n$ distinct scalar fields, and \emph{the wavemap action} is given by the second term in Equation~(\ref{wm1}).

Wavemaps are of special interest for a variety of reasons:
\begin{itemize}
\item They provide a general framework for the combined effects of many coupled scalar fields
\item When $m=2$, $\mathcal{M}$ is a two-dimensional `world sheet' embedded in the target space $\mathcal{N}^{n}$, and the wavemap action (called the Polyakov action in this case) is then conformally invariant for any $n$.
\end{itemize}

The last fact is of importance with relevance to all string actions.

The general description of the evolution of such systems can be considered in the Einstein frame, and in ~\cite{cotsmir01} we showed that for any $m,n$, if we set
\be
\tilde{g}_{\mu\nu}=A(\phi)g_{\mu\nu},
\ee
there exists a conformal equivalence between the general  wavemap-tensor theory~(\ref{wm1})
(with $A,B$ smooth) and the Einstein-wavemap system which is minimally coupled to general relativity derived by the action,
\begin{equation}
    \tilde{S}=\int_{\mathcal{M}}\tilde{L}_{\sigma}dv_{\tilde{g}},\ \ \
    \tilde{L}_{\sigma}=\sqrt{\tilde{g}}( \tilde{R}-\tilde{g}^{\mu\nu}\pi_{ab}\partial_{\mu}\phi^{a}\partial_{\nu}\phi^{b} ).
\end{equation}
where
\begin{equation}
    \pi_{ab}:=\frac{3}{2A^{2}}A_{a}A_{b}+\frac{1}{A}h_{ab}
\end{equation}
for an arbitrary $\mathcal{C}^{\infty}$ function $A$ of $\phi$.

A novel way to realize an inflationary phase in the present context is possible where  the universe inflates exponentially at the critical points of the $00$-component of the wavemap energy-momentum tensor, $T^{00}_WM$. This can be termed $\sigma$\emph{-inflation}, meaning inflation driven both by the coupling $A(\phi)$ and the self-interacting scalar fields $(\phi^{\alpha})$  constitutes a natural way to build a cosmological constant as arising from an earlier state of the universe without using a potential. For a specific analysis of this effect, cf.~\cite{cmt}. 

Finally, wavemaps may be used to provide a new setting for the conformal equivalence theorem. In the following context, a wavemap in the original (Jordan) frame viewed as a matter field there, when conformally transformed becomes a family of scalar fields  on the brane manifold $(\mathcal{M}^{m},g_{\mu\nu})$, while gravity mediates in the bulk $(\mathcal{N}^{n},\pi_{ab})$.

For the sake of illustration, let us assume  that $\textnormal{dim}\mathcal{N}>\textnormal{dim}\mathcal{M}$ so that $\phi(\mathcal{M})\subset\mathcal{N}$ is an $(1+(m-1))$-dimensional subset in the $(1+(m-1)+(n-m))$-dimensional target manifold. Then, a connection between wavemap-tensor theories and braneworlds is revealed: taking a Randall-Sundrum-type brane $(\mathcal{M},\tilde{g})$ with minimal codimension $(n-m=1)$,   the conformal frame  fields are constrained on the brane  $(\mathcal{M}^{m},g_{\mu\nu})$, and their derivatives are taken only with respect to the metric $\tilde{g}$. Additionally, the gravitational field of the Einstein-wavemap system sets freely on the bulk  $(\mathcal{N}^{n},\pi_{ab})$, where the  metric is now given by the $\pi_{ab}$ field. For more details, we refer the reader to Ref.~\cite{cotsmir01}.

Let us end this Section with a brief discussion of another way to represent $f(R)$ or Brans-Dicke-type theories. As we discussed above, a conformal transformation of the form~(\ref{conf}) expresses the general $f(R)$ theory as general relativity plus a self-interacting scalar field defined as in Equation~(\ref{phidefn}) with a potential energy given by~(\ref{pot}).

Independently of the conformal representation of  $f(R)$-gravity~\cite{B-C88}, Brans-Dicke theory~\cite{d2}, or string-type equations~\cite{gasp}, it was found that $f(R)$-gravity can be also expressed alternatively as  a kind of scalar-tensor theory using the Legendre transformation of the lagrangian function in Equation~(\ref{fRLagran}). This can happen only when we restrict the lagrangian function $f(R)$ to be a convex function, i.e.,  $f''(R)\neq 0$, as required by the  general theory of the Legendre transformation, cf.~\cite{ar}.

In that case, we can define the Legendre transformation by  $(R,f(R))\rightarrow(\phi,W(\phi))$, as in ~\cite{ar}, namely,
\be\label{leg}
\phi=f'(R),\quad W(\phi)=Rf'(R)-f(R).
\ee

Here the function  $W(\phi)$,  the Legendre transform of $f(R)$, is not the conformal potential~(\ref{pot}), but it is necessarily convex and involutive.
 That is, if $h$ is the Legendre transform $w$, $w\rightarrow h$, then $w$ will  be  the Legendre transform of $h$. We note that if one assumes that $\phi=R$, then this fixes $f$ to be the quadratic theory $R^2/2+\lambda$. In that case, the quadratic function and its Legendre transform $W$ coincide at corresponding points. We also note that both $f(R)$ and its Legendre transform are `dual' functions to each other, while the spacetime metric is not transformed under the Legendre transformation. A Legendre transformation is not conformal.
Using Equation~(\ref{leg}),  the original $f(R)$ lagrangian can be written as
\be
f(R)=\phi R-W(\phi),
\ee
and this representation is sometimes taken to imply the presence of  a Brans-Dicke-\emph{like} lagrangian theory without the kinetic term of the `scalar field' $\phi$ now defined by~(\ref{leg}), see, e.g.,~\cite{ha,tt,fra1,jk,fra2} (and in many other  references therein, cf.~\cite{sot}). Such a lagrangian has a somewhat restricted physical outlook both as an alternative geometric extension of general relativity, or as a suitable particle physics model. This is  because no `physical' scalar field stress-energy tensor can be formed without a kinetic term if it is to be interpreted as part of the gravitational field, and also because it has a zero Brans-Dicke parameter $\omega$,~\cite{d1}. Nevertheless, this representation of $f(R)$ theory is sometimes considered a suitable, alternative model in scales comparable to the solar system~\cite{ha}.

\section{Slice Energy, Singularities, Multiverse Models}\label{sec3}
We have seen  that the conformal potential~(\ref{pot}) of the scalar field, introduced by the transition from the Jordan to the Einstein frame, appears when the Jordan frame theory is considered in a vacuum. In that case, the scalar field in the Einstein frame is to be regarded as a matter field with potential energy given by~(\ref{pot}).  If we have some other matter field, for example another fluid, in either one of the two frames, then we expect the scalar field to couple with that matter field differently in each frame.

This is of course a well-known issue associated with the conformal transformation, sometimes known as the physicality problem. The physicality problem refers to whether or not there is a way to single out the true, `physical' metric or frame, i.e., the one  satisfying the principle of equivalence (i.e., free-falling particles follow geodesics in spacetime defined by that frame), from another physically inequivalent one between  the two conformally related metric fields~\cite{brans,fra3,cot93,cot95} (see~\cite{fara} for a good review  containing many related references).

Whereas it was originally thought that by giving suitable conservation laws one can arrive at a physically chosen frame, either the Jordan frame~\cite{brans} or the Einstein frame~\cite{fra3}, later it was pointed out by the construction of an explicit counterexample ~\cite{cot93,cot95} that under certain conditions of the scalar field, the Einstein frame is singled out as the physical frame. That counterexample was considered inconclusive in the literature (see for example refs. in ~\cite{cot95}) mainly because of its reliance on the validity of the Bochner's theorem for spacetimes. In fact, a spacetime version of Bochner's theorem was shown to be a valid statement a few years later, cf.~\cite{ro1,ro2,ro3}. However, even if the Einstein frame becomes the natural one for the types of spacetime assumed in the theorem of~\cite{cot95}, it is not known whether or not this conclusion holds for other spacetimes. For other approaches to proving that the Einstein frame is the physical one, see~\cite{ms94}, while for various other results to related cosmological problems in the two frames, the reader is referred to~\cite{od3,od4,od5,od6}.

An alternative way to  study the potential~(\ref{pot}) is to consider its possible effects on standard matter fields  present in the universe. For any matter source and  the frame on which that matter is introduced or of the details of how the scalar field couples to matter,  a matter-scalar field coupling describes their interaction and this will be accompanied by an exchange of energy between the scalar field $\phi$ and the matter component  $\psi$ (or $\tilde{\psi}$ in the conformal Einstein frame). Studying this exchange of energy may provide an alternative way to  the original physicality problem, and may also offer some clues not obtainable otherwise.

The general problem was treated in Ref.~\cite{cots08} by consideration of a new factor, the slice energy with respect to a causal vector field on a space-like hypersurface~\cite{ycb}. This is defined by
the integral (when it exists),
\be
E_{t}=\int_{\mathcal{M}_{t}}P^{a}n_{a}d\m_{t},
\ee
where $n_a$ is the unit normal to $\mathcal{M}_{t}$ in spacetime and $d\m_{t}$ is the volume
element with respect to the spatial metric. The product $P^{a}n_{a}$ is the energy density associated with the energy-momentum vector $P^{a}$ of the stress tensor $T_{\a\b}$ relative to the causal vector field $X_a$, defined as
$P^{b}=X_{a}T^{ab}$.

The general analysis of the problem depends on a number of factors, namely,
\begin{itemize}
\item The  existence of a Killing vector field along the flow lines
\item The  kinematical fluid quantities
\item The positivity of the conformal potential.
\end{itemize}

A general result found in ~\cite{cots08} is that if the slice energy is not conserved during the evolution, then in the Einstein frame free-falling matter will follow geodesics only under very stringent conditions.

Although the general problem is not completely resolved, in some  particular cases one can calculate the exact form of the slice energy in the conformal frame. For instance, for  the case of dust with velocity $\tilde{V}$ tangent to the dust timelines, the field equations are given by,
\begin{equation}
f'R_{\alpha\beta}-\frac{1}{2}g_{\alpha\beta}f-\nabla_{\alpha}\nabla_{\beta}f'+g_{\alpha\beta}\Box_{g}f'=\rho V_{\alpha}V_{\beta},
\end{equation}
and we find that the  total slice energy with respect to the time-like vector field $\tilde{V}$ is in this case given by,
\begin{equation}
    E_{t}(\phi+\textnormal{dust})=E_{0}(\phi+\textnormal{dust})+ \frac{1}{2} \int_{t_{0}}^{t_{1}} \int_{\mathcal{M}_{t}}^{} \tilde{T}^{\alpha\beta}(\phi)(\tilde{\nabla}_{\alpha}\tilde{V}_{\beta}+\tilde{\nabla}_{\beta}\tilde{V}_{\alpha}) d\tilde{\mu}.
\end{equation}

In particular, the total energy of the scalar field-dust system is not conserved in a non-stationary spacetime ($\tilde{V}$ is not a Killing vector field of $\tilde{g}$).

In the more general case of a perfect fluid, an analogous expression exists~\cite{cots08}, namely

\begin{equation}
    E_{t_{1}}(\phi+\textnormal{fluid})=E_{t_{0}}(\phi+\textnormal{fluid})+ \frac{1}{2} \int_{t_{0}}^{t_{1}} \int_{\mathcal{M}_{t}}^{} \tilde{T}^{\alpha\beta}(\phi)(\tilde{\nabla}_{\alpha}\tilde{V}_{\beta}+\tilde{\nabla}_{\beta}\tilde{V}_{\alpha}) d\tilde{\mu} - \int_{t_{0}}^{t_{1}} \int_{\mathcal{M}_{t}}^{} \tilde{p}\tilde{\nabla}^{\alpha}\tilde{V}_{\alpha} d\tilde{\mu}.
\end{equation}

From these equations one may conclude that the slice  energy is generally not conserved as the system evolves over time. This is equivalent to a violation of the principle of equivalence in the Einstein frame. The whole problem deserves further consideration.

Let us finish this Section with a short discussion of how the conformal potential controls the existence  of cosmological singularities.  A sufficient condition for the occurrence of
finite-time singularities is
\be
R_{ab}k^a k^b>0,
\ee
for every time-like vector $k^a$. Combining this with the energy-momentum tensor through the Einstein equations
in $D$ dimensions, we find that the following strong-energy condition is to be satisfied:
\be
(T_{ab}-Tg_{ab}(D-2)^{-1}) k^a k^b>0.
\ee

As was shown in~\cite{B-C88} the above energy condition is not always satisfied in
$f(R)$ gravity theories that are conformally equivalent
to general relativity plus a scalar field with a potential $V(\phi)>0$, so in this case one may draw no conclusions about possible singularities in the Jordan frame of $f(R)$ theories.

However, with $V(\phi)\leq 0$ the above condition is fulfilled and we are led to evolution leading to spacetime  singularities in the case of matter fields satisfying the same condition.

In fact, the case of a positive conformal potential leads to a multiverse model created by quantum fluctuations of the scalar field generated by the conformal transformation, the so-called \emph{quiescent multiverse} developed in Refs.~\cite{c90,B-C91,c94}. This is formally based on a cubic theory of gravity in four spacetime dimensions, but it is actually valid for any polynomial lagrangian of degree $n$ in dimension $D<2n$.

A basic feature of this model which makes the quiescent multiverse distinct from similar models (like the quantum cosmological model of Vilenkin's~\cite{v83}, or Linde's chaotic self-regenerating inflationary universe~\cite{l86}), is the preference of initial scalar field conditions of the form $\phi_0<O(1)$, so that no mini-universe bubbles can collapse to spacetime singularities of infinite density. This is shown in~\cite{c90,B-C91}, where the time it takes for the scalar field (in the conformal frame of the cubic theory) to become infinite decays is
\be
t\sim e^{-\lambda (D,n)\,\phi_0},
\ee
where $\lambda$ is the exponential scalar field potential coupling to $\phi$, so that for small initial values of the scalar field this time is long enough for inflation to take place. (We note that in other multiverse models, this time is vanishingly small). In fact, a more systematic analysis of the quiescent multiverse was given in Ref.~\cite{c94} based on  the behaviour of solutions of a cosmological Fokker-Planck equation, where it was shown that a quiescent multiverse model is a solution of an Ornstein-Uhlenbeck type of stochastic process for the scalar field potential of the cubic lagrangian.

\section{The Traceless Extension}\label{sec4}
Elsewhere we have proposed that the conformal equivalence of $f(R)$ theories with general relativity can be used to support a novel approach to the cosmological constant problem~\cite{notrace1}. In this approach,  \emph{no-scale gravity} is introduced and used to show that  the cosmological constant and the energy  of the vacuum  are two unrelated arbitrary constants. We shall review some of this work in the present Section.

The cosmological constant was introduced by Einstein in 1917 in his fundamental work on  the formulation of modern cosmology, where he altered his original equations of general relativity to better suit the peculiarities of the universe in the largest scales~\cite{ein0}. The fact that the cosmological constant $\Lambda $ has a nature extraneous to both general relativity and quantum field theory follows from the well-studied issue that it leads to the scale $\Lambda=(8\pi G)^{-1/2}$. This implies an observed  discrepancy between the  energy
density of empty space  measured to be around $10^{-47}$ \text{GeV}$^{4}$ calculated using general relativity, and the  vacuum field energy $\Lambda ^{4}/16\pi ^{2}\approx 2\times 10^{71}$ \text{GeV}$^{4}$ using a standard quantum
field theory approximation. This is  the `cosmological constant problem'~\cite{BT, weinberg89}.

In~\cite{notrace1} we suggested that new physics is required to resolve the cosmological constant problem, one that becomes manifest at a hyper-classical, truly cosmological  scale, and contains effects and phenomena not reducible to
either the classical domain of general relativity or the microscopic realm of quantum fields, such as those appearing across the  various \textit{presently}
causally disconnected regions.
The speculative ideas of the multiverse and the anthropic principle~\cite{BT} contain
some of the spirit of the  work in~\cite{notrace1}, but our starting point and
implementation are novel and entirely different.

The above comments lead us to suspect that a theory operating on a cosmological scale must be scale invariant. However, the  standard $f(R)$ theory potential in the conformal representation~(\ref{pot}) breaks scale invariance.
The next  obvious choice, namely, the traceless, `unimodular' Einstein's theory~\cite{el1}, leads in the conformal frame to a dimensionally homogeneous theory with a purely exponential potential~\cite{notrace1}. However, such potentials are ruled out
by the fact that  they lead to no graceful exit from
inflation, and  also because of  disagreements with current observations,~\cite{wein2}, Section 10.2.

However, the conformal frame representation  of the traceless extension of $f(R)$ theory reads~\cite{notrace1},
\begin{equation}
\tilde{G}_{\mu \nu }=\tilde{T}_{\phi ,\mu \nu },  \label{form1}
\end{equation}%
where
\begin{equation}
\tilde{T}_{\phi ,\mu \nu }=\frac{1}{4}(D-1)(D-2)\nabla _{\mu }\phi \nabla
_{\nu }\phi -\frac{1}{2}\tilde{g}_{\mu \nu }\left( \frac{(D-1)(D-2)}{4}%
(\nabla \phi )^{2}+2V(\phi )\right) ,  \label{phi1}
\end{equation}%
with the effective scalar field potential given by,
\begin{equation}
V(\phi )=\left( \frac{D-2}{4D}\right) e^{-\phi }R.  \label{V}
\end{equation}%

These equations are scale invariant, with a potential having a number of novel properties:
\begin{itemize}
\item It is different from both the conformal potential~(\ref{pot}) and that of the conformal representation of the traceless Einstein theory
\item It contains the no-longer-constant (as in traceless GR) scalar curvature $R$, satisfying the constraint equation, which is a second-order differential equation
    \item It is scale invariant on the variables  $(g,f^{\prime })$
\end{itemize}

This no-scale gravity theory has a number of important implications for the cosmological constant problem as well as the interpretation of current observations. We are allowed  to perform  arbitrary linear transformations $\phi\rightarrow a\phi +b$ and shifts
in the origin of the potential without affecting the scale invariance of the field equations. For example,  the no-scale version of the theory
\begin{equation}
f(R)=R+AR^{n},\quad n\geq 2,  \label{dn}
\end{equation}%
when $D=4, n=2$, instead of the standard quadratic potential
proportional to the combination $(1-e^{-\phi})^2$, leads to the \emph{exact} scale invariant potential,
\begin{equation}
V_2=a(1-g e^{-b\phi})+d ,  \label{V2a}
\end{equation}
where $a, b, d, g$ are arbitrary constants. This  satisfies $\lim_{\phi\rightarrow\infty}V_2=a+d$, and therefore we conclude that the cosmological constant becomes an arbitrary constant; the vacuum has energy $V_2(0)$, which is also an arbitrary constant, and the
inflationary plateau is at another, unrelated arbitrary value, $a+d$.

This then provides an unconventional explanation of the cosmological constant. Similarly to an arbitrary constant present in the solution set of a differential equation with its possible values forming the set of all
solutions of the equation (i.e., the general solution),  the  possible values of the cosmological constant are distributed on all different local mini-universes correlated in the so-formed multiverse.

The explanation of the cosmological constant values in the context of the no-scale theory described above leads to a set of different domains, one of which is our local mini-region with its observed value of that constant. It means that the cosmological constant takes randomly distributed values
in any causally connected  domain in the  landscape and  the same holds for
the other two parameters of the no-scale theory, $V(0)$ and the asymptotic potential
value $V_\infty=a+d$. For more results of this on-going program, see~\cite{notrace1}.

\section{Palatini Variation}\label{sec5}

In the search of a possible geometric extension  of general relativity we consider
an arbitrary connection $\nabla$  totally unrelated to
the metric, i.e., $\nabla\mathbf{g}\neq0,$, the connection coefficients
$\Gamma_{bc}^{a}$, are independent functions of the metric
components $g_{ab}$. The associated variational method is the
Palatini variation and consists of  independent variations of the metric and the connection.

An interesting property of the Palatini variation is that for lagrangians that
are functions of curvature invariants we obtain second order field equations.
For simplicity, we restrict ourselves to lagrangians that are smooth functions
of the scalar curvature $R,$
\begin{equation}
L=wf\left(  R\right)  ,\ \ \ \ \ \ \ \ \ w:=\sqrt{-g}.\label{lafr}%
\end{equation}

For the case of combinations of the curvature invariants  $R_{ab}R^{ab}$ and
$R_{abcd}R^{abcd}$, see~\cite{cotsmirque99}. The Ricci tensor depends on the connection $\Gamma$ and its derivatives, and it  is independent of the metric. Variation of the
\begin{equation}
S=\int wd^{4} x f\left(  R\right)  \label{afr}%
\end{equation}
with respect to the metric and the connection yields, respectively,
\begin{equation}
f^{\prime}R_{\left(  ab\right)  }-\frac{1}{2}fg_{ab}=0,\label{pfr1}%
\end{equation}%
\begin{equation}
\nabla_{a}\left(  wf^{\prime}g^{bc}{}\right)  =0.\label{pfr2a}%
\end{equation}

Equation~(\ref{pfr2a}) implies that
\begin{equation}
\partial_{a}\left(  f^{\prime}g_{bc}\right)  =\Gamma_{ba}^{m}f^{\prime}%
g_{mc}+\Gamma_{ca}^{m}f^{\prime}g_{mb},\label{pm}%
\end{equation}
i.e., the covariant derivative of $f^{\prime}g_{ab}$ with respect to the
connection $\Gamma$ vanishes. Therefore, the connection $\Gamma$ is the
Levi-Civita connection for the metric $\widetilde{g}_{ab}:=f^{\prime}g_{ab}$.

We now turn our attention to~(\ref{pfr1}). Taking the trace of~(\ref{pfr1}) we
find
\begin{equation}
f^{\prime}\left(  R\right)  R=2f\left(  R\right)  , \label{tracfr}%
\end{equation}
and integrating we obtain
\begin{equation}
f\left(  R\right)  =R^{2}, \label{quadr}%
\end{equation}
up to a constant factor. Accordingly,  Equation~(\ref{pfr1}) becomes,
\begin{equation}
R_{ab}-\frac{1}{4}Rg_{ab}=0, \label{pfr1a}%
\end{equation}
provided that $f^{\prime}\left(  R\right)  \neq0.$ The scalar curvature is undetermined because~(\ref{pfr1a}) is traceless, but adding a traceless energy momentum tensor, that is a traceless matter lagrangian in the action~(\ref{afr}), Equation~(\ref{pfr1a}) becomes identical to  Einstein's traceless relativity theory~\cite{ein3,notrace1,weinberg89}. Moreover,  $R_{ab}$ is invariant
under the transformation $g_{ab}\rightarrow \tilde{g}_{ab}=f^{\prime}g_{ab}=$ (which using Equation~(\ref{tracfr}) can be written as $\tilde{g}_{ab}=k^2 R g_{ab}$, $k^2$ a constant), i.e.,
$\widetilde{R}_{ab}=R_{ab}.$ Hence, the field Equation~(\ref{pfr1a}) becomes
\begin{equation}
\widetilde{R}_{ab}-\frac{1}{4}\tilde{R}\tilde{g}_{ab}=0. \label{confr1}%
\end{equation}

Therefore,  the Palatini theory~(\ref{confr1}), or, equivalently,  the lagrangian~(\ref{afr}) (also with traceless matter added), is conformally invariant  with constant conformal  scalar curvature
$\tilde{R}=k^{-2}.$ This is in sharp contrast to the corresponding situation in the conformal frame of the standard Einstein traceless relativity~\cite{notrace1}.

From~(\ref{quadr}) one could draw the conclusion that the Palatini method does
not apply to general lagrangians $f\left(  R\right)  $ as it forces them  to
be strictly quadratic. However, for an arbitrary
smooth function $f$, Equation~(\ref{tracfr}) is also  an algebraic
equation for $R$,~\cite{ffv}. If the  roots of this equation are denoted by
$\rho_{1},\rho_{2},\dots,$ then we obtain a whole series of conformally invariant theories,
each having  constant scalar curvature. 

In the following we shall argue that a less restrictive way for independently
varying the metric and  connection fields is  adding a compatibility
condition for the metric and connection and introducing Lagrange multipliers, a method which is sometimes referred to in the literature as  the \emph{constrained Palatini variation} (CPV).

A general higher-order lagrangian with arbitrary  matter
couplings is given by the form,
\begin{equation}
L\left(  g,\nabla g,...,\nabla^{\left(  m\right)  }g;\,\psi,\nabla
\psi,...,\nabla^{\left(  p\right)  }\psi\right)  . \label{CPLag}%
\end{equation}

For an arbitrary symmetric connection $\Gamma_{bc}^{a}$, we have
\begin{equation}
\Gamma_{ab}^{c}=\left\{  _{ab}^{c}\right\}  -\frac{1}{2}g^{cm}\left(
\nabla_{b}g_{am}+\nabla_{a}g_{mb}-\nabla_{m}g_{ab}\right)  . \label{ident1}%
\end{equation}
which, restricted to  Weyl geometry,
$\nabla_{c}g_{ab}=-Q_{c}g_{ab},$ with $Q_{c}$  a covariant vector
field,  becomes
\begin{equation}
\Gamma_{ab}^{c}=\left\{  _{ab}^{c}\right\}  +\frac{1}{2}g^{cm}\left(
Q_{b}g_{am}+Q_{a}g_{mb}-Q_{m}g_{ab}\right)  . \label{ident2}%
\end{equation}

We then introduce the difference tensor,
\begin{equation}
C_{ab}^{c}=\Gamma_{ab}^{c}-\left\{  _{ab}^{c}\right\}  , \label{tensC}%
\end{equation}
The constrained Palatini variation is affected  by adding the following term as a constraint (with Lagrange
multiplier $\Lambda$) to the original lagrangian~(\ref{CPLag}), namely,
\begin{equation}
L_{c}\left(  g,\Gamma,\Lambda\right)  =\Lambda_{r}^{\ mn}\left[  \Gamma
_{mn}^{r}-\left\{  _{mn}^{r}\right\}  -C_{mn}^{r}\right]  . \label{const}%
\end{equation}

In Riemannian geometry, we have
\[
L_{c}\left(  g,\Gamma,\Lambda\right)  =\Lambda_{r}^{\ mn}\left[  \Gamma
_{mn}^{r}-\left\{  _{mn}^{r}\right\}  \right]  .
\]

Then ~(\ref{CPLag}) becomes,
\[
L=L\left(  g;\,\Gamma,\nabla\Gamma,...,\nabla^{\left(  m-1\right)  }%
\Gamma;\,\psi,\nabla\psi,...,\nabla^{\left(  p\right)  }\psi\right)
\]
and an independent variation of the   action
\begin{equation}
S=\int w d^{4} x\left[  L\left(  g,\Gamma,\psi\right)  +L_{c}\left(  g,\Gamma
,\Lambda\right)  \right]  \label{CPAct}%
\end{equation}
with respect to the fields $g$, $\Gamma$, $\Lambda$, and $\psi$, implies that
the  field equations, coming from the general lagrangian
(\ref{CPLag}) upon the standard Einstein- Hilbert variation, are identical to those
obtained from the  `constrained' lagrangian,
\begin{equation}
L^{\prime}\left(  g,\Gamma,\Lambda,\psi\right)  =L\left(  g,\Gamma
,\psi\right)  +L_{c}\left(  \Lambda,\Gamma\right)  \label{CPCLag}%
\end{equation}
using  the constrained Palatini variation.

As an application, consider the
constraint~(\ref{const}) in Weyl geometry,
\begin{equation}
L_{c}\left(  \Lambda,C\right)  =\Lambda_{c}^{ab}\left[  \Gamma_{ab}%
^{c}-\left\{  _{ab}^{c}\right\}  -\frac{1}{2}g^{cm}\left(  Q_{a}g_{mb}%
+Q_{b}g_{am}-Q_{m}g_{ab}\right)  \right]  .\label{wcons}%
\end{equation}

Let us apply the CPV method for the case of the lagrangian $L=f\left(R\right).$ Variation of
the Lagrange multiplier recovers Equation~(\ref{ident2}) of the Weyl
connection, whereas varying  the connection yields the explicit form
of the Lagrange multiplier. Finally, the metric variation  gives
 (see~\cite{cotsmirque99} for details),
\begin{equation}
f^{\prime}R_{\left(  ab\right)  }-\frac{1}{2}fg_{ab}-\nabla_{a}\nabla
_{b}f^{\prime}+g_{ab}\square f^{\prime}=M_{ab},\label{Wfe}%
\end{equation}
where `matter part' $M_{ab}$ is given by
\begin{equation}
M_{ab}=-2Q_{(a}\nabla_{b)}f^{\prime}-f^{\prime}\nabla_{\left(  a\right.
}Q_{\left.  b\right)  }+f^{\prime}Q_{a}Q_{b}+g_{ab}\left(  2Q_{m}\nabla
^{m}f^{\prime}-Q^{2}f^{\prime}+f^{\prime}\nabla^{m}Q_{m}\right)
.\label{Mtens}%
\end{equation}

The usual Riemannian geometry field equations from a Hilbert variation correspond to the degenerate case $Q_{a}=0$,
\[
f^{\prime}R_{ab}-\frac{1}{2}fg_{ab}-\nabla_{a}\nabla_{b}f^{\prime}%
+g_{ab}\square f^{\prime}=0,
\]
which are conformally dynamically
equivalent to Einstein equations with a self-interacting scalar field as shown earlier. We can generalize this result for the present case and note that the Weyl vector transforms as
\[
\widetilde{Q}_{a}=Q_{a}-\nabla_{a}\ln f^{\prime},
\]
giving in the Einstein frame the result
\begin{equation}
\widetilde{G}_{ab}=\widetilde{M}_{ab}^{Q}-\widetilde{g}_{ab}V\left(
\phi\right)  ,\label{WCeqn}%
\end{equation}
where
\[
\widetilde{G}_{ab}:=\widetilde{R}_{\left(  ab\right)  }-\frac{1}{2}%
\widetilde{R}\widetilde{g}_{ab}%
\]
and
\[
\widetilde{M}_{ab}^{Q}:=-\widetilde{\nabla}_{\left(  a\right.  }\widetilde
{Q}_{\left.  b\right)  }+\widetilde{Q}_{a}\widetilde{Q}_{b}+\widetilde{g}%
_{ab}\left(  -\widetilde{Q}^{2}+\widetilde{\nabla}^{m}\widetilde{Q}%
_{m}\right) ,
\]
and the potential takes its usual form~(\ref{pot}). When $\widetilde{Q}_{a}=0$ (`the Riemannian case') and the Weyl vector is a gradient, $Q_{a}=\nabla_{a}\Phi$,  the field Equation~(\ref{WCeqn}), are the Einstein equations with a
cosmological term and a source term $\widetilde{M}_{ab}^{Q}$, and we are back to the Palatini variation of the lagrangian $L=f\left(  R\right).$ 
We therefore conclude that the Palatini variation method cannot deal with
general Weyl geometry in a satisfactory manner. For more applications of
Palatini-type theories, see~\cite{capo1,olmo1,cope2,olmo2} and refs. therein.

\section{Conformal Raychaudhuri Equation}\label{sec6}
\label{corayc} In this and the next Section, we discuss the question of whether the present
isotropic state of the universe can be obtained from arbitrary initial
conditions in higher-order gravity theories. In the present Section, we prove
that the generalized Raychaudhuri equation is conformally equivalent to the
usual one met in general relativity but coupled to the wave
equation satisfied by the conformal scalar field. In the next Section, we discuss the validity of  the cosmic no-hair conjecture for the Bianchi models in the Einstein frame of the $R+\alpha
R^{2}+L_{\mathrm{matter}}$ theory.

For the theory,
\begin{equation}
L=f(R)+L_{m},\label{lama}%
\end{equation}
on a Bianchi spacetime, let
$\mathbf{n}=\partial/\partial t$ be the hypersurface orthogonal unit vector
field, and $h_{ab}=g_{ab}+n_{a}n_{b},$ be the spatial metric related to the spacetime metric. The extrinsic curvature $K_{ab}$ of the
spacelike hypersurfaces $K_{ab}=\nabla_{a}n_{b}$ splits into its trace and traceless parts
$K=h^{ab}K_{ab}$ $\sigma_{ab}$, respectively, as follows,
\begin{equation}
K_{ab}=\frac{1}{3}Kh_{ab}+\sigma_{ab}.\label{excu4}%
\end{equation}

Since for a congruence of timelike geodesics hypersurface orthogonal the
rotation $\omega_{ab}$ vanishes, the Raychaudhuri equation takes the form
\cite{he}
\begin{equation}
\frac{dK}{dt}=-\frac{1}{3}K^{2}-2\sigma^{2}-R_{ab}n^{a}n^{b}. \label{ra}%
\end{equation}

The Raychaudhuri equation is of a a geometric nature and has nothing to do with
the field equations that hold in a given spacetime, however, since  the Ricci tensor in the last term satisfies the $f\left(  R\right)  $
Equation~(\ref{fRVeqns}) with matter, it is not linearly coupled to
it. As a result, the positivity of the term $R_{ab}n^{a}n^{b}$ is not
guaranteed by imposing the strong energy condition (SEC), i.e., $T_{ab}%
n^{a}n^{b}+\frac{1}{2}T\geq0$, on the stress-energy tensor. The equivalence
between the timelike convergence condition (TCC), i.e., $R_{ab}n^{a}n^{b}\geq
0$, and the SEC encountered in general relativity, no longer holds in $f\left(
R\right)  $ theories. 

We now show how the relevant quantities appearing in~(\ref{ra}) transform
under the conformal transformation~(\ref{phidefn}). We note that the
congruence of timelike geodesics remains timelike since $\left(
M,\mathbf{g}\right)  $ and $\left(  M,\tilde{\mathbf{g}}\right)  $ have
identical causal structure. The congruence also remains hypersurface
orthogonal (hence $\tilde{\omega}_{ab}=0$) in the conformal manifold
$(M,\mathbf{\tilde{g}})$ since a conformal transformation preserves angles. We
define the unit (with respect to the metric $\tilde{\mathbf{g}}$) vector field
tangent to the associated congruence of timelike curves by $\tilde{n}%
^{a}=e^{-\phi/2}n^{a}$ (as a consequence the time derivative operator
in $(M,\tilde{\mathbf{g}})$ becomes $d/d\tilde{t}=e^{-\phi/2}d/dt$) and
$\tilde{h}_{ab}=\exp(\phi)h_{ab}$. We can now define the extrinsic curvature,
shear, expansion, etc. in the spacetime $(M,\mathbf{\tilde{g}}),$ and the
associated quantities become
\begin{align}
\tilde{K}_{ab} &  =h_{a}^{c}h_{b}^{d}\tilde{\nabla}_{(c}\tilde{n}%
_{d)},\ \ \ \ \ \tilde{\sigma}_{ab}=\tilde{K}_{ab}-\frac{1}{3}\tilde{K}%
\tilde{h}_{ab},\nonumber\\
\tilde{K} &  =\tilde{h}^{ab}\tilde{K}_{ab}=\tilde{\nabla}_{a}\tilde{n}%
^{a},\ \ \ \ \ \dot{\tilde{n}}^{a}=\tilde{n}^{c}\tilde{\nabla}_{c}\tilde
{n}^{a}.\label{acce}%
\end{align}

The acceleration term appearing in~(\ref{acce}) is in general non-zero since
our curves are no longer geodesics. The transformed Raychaudhuri Equation
(\ref{ra}) is then given by
\begin{equation}
\dot{\tilde{K}}:=\tilde{n}^{a}\tilde{\nabla}_{a}\tilde{K}=-\frac{1}{3}%
\tilde{K}^{2}-2\tilde{\sigma}^{2}-\tilde{R}_{ab}\tilde{n}^{a}\tilde{n}%
^{b}+\tilde{\nabla}_{a}\dot{\tilde{n}}^{a}.\label{cora}%
\end{equation}

For the computation of the tensor field $\tilde{\nabla}_{(c}\tilde{n}_{d)}$ we
employ the relation between $\tilde{\nabla}_{a}\omega_{b}$ and $\nabla
_{a}\omega_{b}$ in terms of the conformal factor $\Omega^{2}=\exp\phi.$ This
allows the computation of $\tilde{K}^{2},$ $\tilde{\sigma}^{2}$ and
$\dot{\tilde{n}}^{a}.$ After a lengthy algebraic calculation, Equation~(\ref{cora}) becomes,
\begin{align}
\dot{\tilde{K}}:=\tilde{n}^{a}\tilde{\nabla}_{a}\tilde{K} &  =-\frac{1}%
{3}\tilde{K}^{2}-2\tilde{\sigma}^{2}-\tilde{R}_{ab}\tilde{n}^{a}\tilde{n}%
^{b},\label{syra}\\
\ddot{\tilde{\phi}}+\tilde{K}\dot{\tilde{\phi}}+\frac{dV_{\ast}}{d\phi} &
=0,\quad\quad V_{\ast}(\phi)=\frac{2}{3}V(\phi).\label{syfi}%
\end{align}

Using this system we conclude that Wald's no-hair
theorem~\cite{Wald83} also holds for $f(R)$ gravity with a flat potential (see next Section). Also,
the Barrow-Ottewill~\cite{C-F93} quasi-exponential solution $\exp(Bt-At^{2})$ of the theory $f(R)=R+\alpha R^{2}$ is a
stable attractor of all isotropic solutions of~(\ref{lama}). Apart from these results, for an exponential
potential $V$, initial expanding
Bianchi models are attracted by the isotropic power-law inflationary solution
leading to double inflation in the Jordan frame.

\section{Cosmic No-Hair Theorem in Quadratic Gravity}\label{sec7}
Although inflation provides an attractive explanation of the homogeneity and
isotropy of the universe,  it is not obvious whether
initial inhomogeneities and anisotropies during inflation will be smoothed out eventually. With
regard to this issue, Gibbons and Hawking~\cite{giha} and Hawking and Moss
\cite{hamo} proposed the cosmic no-hair conjecture, which states that \emph{All
expanding models with a positive cosmological constant
approach the de Sitter solution asymptotically.
}Wald~\cite{Wald83} proved that this
conjecture is true for Bianchi, and his proof has become
a standard approach  for later works~\cite{jess}. In this section we prove the no-hair theorem in a curvature-squared theory for
all Bianchi universes with matter  satisfying suitable energy conditions. 

For such a theory,  the conformal frame  field equations
are Einstein equations with a self-coupled scalar field on $\left(  M,\tilde
{\mathbf{g}}\right)  $
\begin{equation}
\tilde{G}_{ab}=\tilde{T}_{ab}^{\phi}+\tilde{T}_{ab},\ \ \ \ \ \ \ \ \tilde
{\square}\phi-V^{\prime}\left(  \phi\right)  =0,\label{box}%
\end{equation}
with potential $V$ given by Equation~(\ref{pot}),
\begin{equation}
V=\frac{1}{8\alpha}\left[  1-\exp\left(  -\sqrt{2/3}\phi\right)  \right]
^{2}.\label{pote5}%
\end{equation}

This potential  is almost constant on the plateau, $V_{\infty}:=\lim_{\phi
\rightarrow+\infty}V\left(  \phi\right)  =1/\left(  8\alpha\right)$, so that $V_{\infty}$ behaves as a
cosmological term, allowing for inflation.

For  simplicity we drop the tilde. Then, the $00$-component of the field equations is
\begin{equation}
\frac{1}{3}K^{2}=\sigma^{2}+\rho+\frac{1}{2}\dot{\phi}^{2}+V-\frac{1}%
{2}\;^{\left(  3\right)  }R, \label{ooco5}%
\end{equation}
the Raychaudhuri equation with a scalar field reads
\begin{equation}
\dot{K}=-\frac{1}{3}K^{2}-2\sigma^{2}-\left(  \rho+3p\right)  -\dot{\phi}%
^{2}+V, \label{rayc5}%
\end{equation}
and the scalar field equation, the second equation in ~(\ref{box}), becomes
\begin{equation}
\ddot{\phi}+K\dot{\phi}+V^{\prime}\left(  \phi\right)  =0. \label{emsf}%
\end{equation}
In the following, first we show:
\begin{itemize}
\item  all initially expanding Bianchi
models  initially on the flat plateau~(\ref{pote5}), except probably Bianchi IX, with  matter  satisfying the strong and dominant energy conditions, rapidly approach de
Sitter space-time
\item  Bianchi IX universes also
isotropizes if the scalar three-curvature $^{\left(  3\right)  }%
R$ is initially less than the scalar field potential $V$.
\item the time needed for the potential energy to reach its
minimum is much larger than the isotropization time of the models. (Hence  the universe reaches the potential minimum
where $\Lambda=0$ and evolves as in standard Fiedmann cosmology.)
\end{itemize}

We assume that initially  the state of the system evolves  on the flat plateau, i.e., that the initial $\phi_{i}$  value is large and
positive. Inflation lasts during the time interval $t_{f}-t_{i}$ required  for the
scalar field to evolve from  $\phi_{i}$ to a smaller value
$\phi_{f}$ determined by the condition that $V\left(
\phi_{f}\right)  \simeq\eta V_{\infty},$ where $\eta$ is of
order $0.9$. The proof consists in
estimating suitable bounds for the function $S\left(  t\right)  $,
\cite{mosa}, defined by
\[
S=\frac{1}{3}K^{2}-E.
\]

It is shown in~\cite{comi} that for all Bianchi models except type-IX
\begin{equation}
0\leq S\left(  t\right)  \leq\frac{3m^{2}}{\sinh^{2}\left(  mt\right)
},\;\;\,m:=\sqrt{\eta V_{\infty}/3}.\label{upbo}%
\end{equation}

From~(\ref{ooco5}) it follows  that the shear,  three-curvature, and energy
density of  matter rapidly approach zero. Thus, the universe isotropizes
within a time of the order  of $1/\sqrt{V_{\infty}}\sim10^{7}\;t_{PL}.$

For Bianchi IX universes, since  the determinant of the three
metric determines the largest positive value that the
spatial curvature can achieve,  a lower
bound for $S$ can be obtained. It is shown in~\cite{comi} that if initially $V>\,^{\left(  3\right)
}R_{\max}$, then
\begin{equation}
-\frac{1}{2}\,^{\left(  3\right)  }R_{\max}\left(  t_{i}\right)  \exp\left[
-\sqrt{\frac{2}{3}\eta V_{\infty}}\left(  t-t_{i}\right)  \right]  \leq
S\leq\max\left\{  0,\;\frac{3m^{2}}{\sinh^{2}\left(  mt\right)  }\right\}
,\label{upper}%
\end{equation}
and therefore $S$ vanishes almost exponentially. From~(\ref{ooco5}),
$-2S\leq\,^{\left(  3\right)  }R\leq\,^{\left(  3\right)  }R_{\max}$, so
$^{\left(  3\right)  }R$ decays to zero. As in the case of the other Bianchi
cosmologies,  if initially the universe evolves on the plateau, then within a time of the order $\sim1/\sqrt
{V_{\infty}}$ the shear,  scalar
three-curvature, as well as all the stress-energy tensor components,  decay to zero
almost exponentially fast.

Since in our present case the cosmological term
 vanishes asymptotically, the universe may not have enough
time to isotropize during the scalar field evolution. However, we can show that the vacuum energy is not completely exhausted before the universe
isotropizes. If at the beginning of inflation the universe is on the
flat plateau of the potential, then  the time $t_{f}-t_{i}$
 is smaller in the absence of damping than  in its presence. In the former,
$t_{f}-t_{i}$ is more than $65$ times the time $1/\sqrt
{V_{\infty}}$ of isotropization, if one takes
$\phi_{f}$ to be such that $V\left(  \phi_{f}\right)  \simeq\eta V_{\infty}$. On the other hand, damping will increase the interval $t_{f}-t_{i}$, thus giving more inflation sufficient for complete isotropization, QED.

We end this Section with some comments about our method of proof for all the Bianchi universes except IX in a vacuum  compared to that in~\cite{ma88}. In
that case, $^{\left(  3\right)  }R\leq0$ and $\rho=p=0,$ while during inflation $V$ remains less than $V_{\infty}.$ Hence, Wald's method applies without using the scalar field equation, and from  Equations~(\ref{ooco5}) and (\ref{rayc5}) we find
\[
\sqrt{3V_{\infty}}\leq K\leq\dfrac{\sqrt{3V_{\infty}}}{\tanh\mu t}%
,\ \ \sigma^{2}\leq\dfrac{V_{\infty}}{\sinh^{2}\mu t},\ \ \ \dot{\phi}^{2}%
\leq\dfrac{2V_{\infty}}{\sinh^{2}\mu t},
\]
with $\mu=\sqrt{V_{\infty}/3}.$

\section{The Recollapse Problem}\label{sec8}
In this section, it is shown that an initially expanding, closed FRW universe filled with a perfect fluid with an equation of state $p=(\gamma-1)\rho,$ with $2/3<\gamma\leq2$,
starting close to  Minkowski space, cannot avoid recollapse in the Einstein frame of the
$R+\alpha R^{2}+L_{m}$ theory. The field equations are as follows,,
\begin{equation}
\left(  \frac{\dot{a}}{a}\right)  ^{2}+\frac{k}{a^{2}}=\frac{1}{3}\left(
\rho+\frac{1}{2}\dot{\phi}^{2}+V\left(  \phi\right)  \right)  ,\label{fri1jm}%
\end{equation}
\begin{equation}
\frac{\ddot{a}}{a}=-\frac{1}{6}\left(  \left(  3\gamma-2\right)  \rho
+2\dot{\phi}^{2}-2V\right)  ,\label{fri2jm}%
\end{equation}
\begin{equation}
\ddot{\phi}+3\frac{\dot{a}}{a}\dot{\phi}+V^{\prime}\left(  \phi\right)
=0,\label{emsjm}%
\end{equation}
\begin{equation}
\dot{\rho}+3\gamma\rho\frac{\dot{a}}{a}=0,\label{conssfjm}%
\end{equation}
where  $V$  is given by
(\ref{pote5}). Using the constraint~(\ref{fri1jm})
to eliminate $a$,$\,$the change of variable $u:=-e^{-\phi}$  to remove the
transcendental functions, and a rescaling, the evolution Equations
(\ref{fri2jm})--(\ref{conssfjm}) can be written as a four-dimensional
dynamical system in the variables $\left(  u:=e^{-\phi},y:=\dot{\phi}%
,\rho,H:=\dot{a}/a\right)  $:
\begin{align}
\dot{u} &  =-uy,\nonumber\\
\dot{y} &  =-u+u^{2}-3Hy,\nonumber\\
\dot{\rho} &  =-3\gamma\rho H,\nonumber\\
\dot{H} &  =\frac{1}{4}\left(  1-u\right)  ^{2}-\frac{1}{2}y^{2}-\frac
{3\gamma-2}{6}\rho-H^{2},\label{sys3}%
\end{align}
subject to the constraint%
\[
H^{2}-\frac{1}{3}\rho-\frac{1}{4}y^{2}-\frac{1}{4}\left(  1-u\right)  ^{2}<0.
\]

The system has two non-hyperbolic equilibria; therefore, their stability cannot be usefully deduced from linearization.

EQ1:\- $\left(  u=0,\dot{\phi}=0,\rho=0,H=\sqrt{V_{\infty}}\right)  .$ This is a de Sitter space with a cosmological constant equal to
$\sqrt{V_{\infty}}$.

EQ2: $\left(u=1,\dot{\phi}=0,\rho=0,H=0\right).$ This describes end states of ever-expanding universes with $H\rightarrow0$. In such models, the scalar field evolves to the potential minimum while the scale factor diverges.

Applying center manifold theory, it may be shown that the EQ1 equilibrium is unstable even for large
initial $H$-values near the flat plateau, cf.~\cite{miri2}. Further,
using normal form theory, a more detailed study of the
solutions of this model near EQ2 was performed in~\cite{miri2}.
For  $\gamma\in\left[  2/3,2\right]$, initially expanding universes evolve such that $H$ monotonically  decreases,  while the other
variables become oscillatory with ever decreasing amplitudes. At some later time, when the scale factor reaches a maximum,   $H$ passes through zero and continues to decrease while the other
variables oscillate,  now with an ever increasing amplitude. This way one reaches the conclusion  that an initially expanding closed universe in the neighbourhood of the EQ2 solution cannot avoid recollapse.

These results describe the behaviour of the
system near its equilibria and are not sufficient to decide what happens in general. This problem requires a detailed investigation of the global properties of the solutions away from the equilibria.

\section{Discussion}\label{sec9}
In this paper, we have presented critical appraisal of many results  that together comprise the core of an area that may be called \emph{conformal cosmology}--the study and implications of conformal relations in theoretical cosmology. We tabulate below the most important results reviewed here, and include  brief discussions of remaining problems and future directions in conformal cosmology. It is perhaps true that many of the known results in this field are not completely appreciated by the community even now, after many years of their appearance and rapid development.

In the Introduction, we presented  three main subtopics in conformal cosmology of special interest in this paper, namely $f(R)$ theory, Brans-Dicke-type gravity, and
effective string actions, with a special focus on the first, and showed that under the conformal transformation of the metric given by Equation~(\ref{conf}), all three types of gravitational action and associated field equations contain a `basic part' that becomes equivalent to general relativity, plus an enigmatic,  self-interacting scalar field with a particular potential. This holds for   the $f(R)$ lagrangian, various Brans-Dicke-types of theory, and for the gravi-dilaton part of string-theoretic gravities.

In Section~\ref{confthm}, we provided a critical survey of three different aspects of the conformal potential that describes the aforementioned equivalence. These are  basic conformal potential properties, the wavemap-tensor extension, and the Legendre representation. More specifically, firstly, we reviewed four most basic and fundamental properties that this potential has, namely,
\begin{itemize}
\item The conformal factor $\Omega$ is non-singular
\item The scalar field $\phi$ is defined or coupled to the spacetime geometry itself
\item The mathematical conformal equivalence leads to the problem of the physical relation of the two frames, the Jordan and the Einstein frame
\item The conformal potential appears already in vacuum, but a presence of matter fields leads to physical effects not present in general relativity.
\end{itemize}

Secondly, we described the generalisation of the conformal equivalence to an arbitrary number of coupled scalar fields, the so-called wavemap-tensor theory that has an interesting geometric interpretation, and also briefly presented the emergence of inflation and braneworld theory in this context. Thirdly, we discussed an alternative representation that applies to both the $f(R)$ and the Brans-Dicke theories, namely the Legendre transformation. This is sometimes employed in the literature to discuss a relation of $f(R)$ to scalar-tensor theory, although there are limitations of this representation as we showed in the paper.

In Section~\ref{sec3}, we studied a number of topics all using the conformal potential by necessity. These are the physicality issue, the study of slice energy, the existence of singularities in the two frames, and the possibility of a multiverse created by the conformal scalar field's quantum fluctuations. The  physicality problem, still unresolved as of today,  is the issue of whether or not one of the two conformally related metrics can become the true, physical one to use when measuring intervals in spacetime while the other is not, and admits a number of different approaches for its resolution. We showed that one of these, the consideration of the slice energy (the total  observed energy on a hypersurface) of the scalar field and a matter source (say, dust) may not be conserved during evolution due to an exchange of energy between the different components.
The singularity theorems of general relativity can be transferred between frames and this was discussed more fully here. In fact, one may consider a singularity-free solution emerging from the primordial quantum fluctuations of the conformal scalar field and describing a multiverse having `quiescent' characteristics and being distinct from other such models.

In Section~\ref{sec4}, we surveyed the difficulties associated with the Jordan frame field equations (say, of $f(R)$ gravity)  not being scale invariant. In the conformal frame these difficulties show up in the conformal potential being responsible for breaking the scale invariance of the vacuum Einstein equations. We formulated \emph{no-scale gravity},  as a traceless $f(R)$ theory in the conformal frame, and introduce it as a way to remedy this problem. In the resulting setup, the cosmological constant acquires a novel statistical interpretation  as a randomly distributed parameter, and the same is true for the unrelated initial value $V(0)$ and height of the (now scale invariant) conformal potential. It is important that in the no-scale gravity theory, inflation appears quite naturally for all dimensions and conformal potentials with adjustable slow-roll parameters compatible with current data.

In Section \ref{sec5}, we discuss another way to avoid the higher-order terms present in the Jordan frame equations, in particular those of the $f(R)$ gravity: the now popular Palatini method. We show that  for generic higher-order lagrangian theories of gravity the usual variational analysis (independent variation of the metric and connection) is efficient only by the so-called constrained Palatini variation (CPV). Another property of such `Palatini gravities' is their relation to Weyl geometry, a generalized form of Riemannian structure that plays a pivotal role here, as well as the fact that these theories admit a conformal representation by going to the Einstein frame in the usual way. In fact, the conformal structure of Palatini gravity reveals the superiority of the CPV method over the more standard Palatini variation.

Finally, in Sections \ref{sec6}--\ref{sec8} we review basic results related to three fundamental cosmological problems in the two conformal frames:
\begin{itemize}
\item The question of whether the present state in the evolution of the universe could be obtained from generic initial conditions (`why the present universe appears isotropic and homogeneous?')
\item The question of whether a primordial inflationary state is typical in the space of initial states (`expanding universe with a positive cosmological constant approach the de Sitter solution', - the cosmic no-hair conjecture)
\item The closed-universe recollapse conjecture.
\end{itemize}

To achieve the results, we presented in Section \ref{sec6} an analysis of the Landau-Raychaudhuri equation in both conformally related frames for the case of $f(R)$ gravity.

Overall, we can now list a number of outstanding directions of research that stem from the discussion of the results presented in this paper. We believe that research in the topics given below will have a positive effect to the development of the field of conformal cosmology.
\begin{enumerate}
\item Study of physical effects associated to the different ways of matter-scalar field as opposed to matter-geometry couplings in both frames
\item Develop wavemap-tensor cosmology
\item Solve the physicality problem and find its relation to dark energy
\item Study the dependence of the nature of dynamical singularities on the change of frame
\item Develop no-scale cosmology
\item Develop more fully the role of the Legendre transformation in Palatini gravity
\item Develop more fully the issues 1--6 for the string cosmology actions.
\end{enumerate}

\end{document}